
\magnification=1200 \def\wc{\hangindent=4em \hangafter=1 \noindent}
\baselineskip 14pt \parskip 3pt \null 
\headline={\ifnum\pageno=1\hfil\else\hfil\tenrm--\ \folio\ --\hfil\fi}
\footline={\hfil}

\centerline{\bf On the Detectability of Very Massive Compact Objects}
\centerline{\bf with Gravitational Microlensing}
\vskip 0.5cm
\centerline{by}
\vskip 0.5cm
\centerline{B. Paczy\'nski}
\vskip 0.5cm
\centerline{Princeton University Observatory, Peyton Hall, Princeton,
NJ 08544-1001, USA}
\centerline{Visiting Scientist, National Astronomical Observatory, Mitaka,
Tokyo, 181, Japan}
\centerline{e-mail: bp@astro.princeton.edu}

\vskip 0.5cm
\centerline{\it Received: .........................}
\vskip 0.5cm
\centerline {ABSTRACT}
\vskip 0.5cm

If the dark halo of our galaxy is made of compact objects as massive as
$ M \sim 10^6 ~ M_{\odot} $, their detection by means of ordinary
microlensing searches would take a very long time  as the characteristic
time scale of such a lensing event is $ t_0 \sim 200 $ years.  Fortunately,
the very high magnification events of the numerous faint stars, which are
normally well below the detection threshold, have short duration peaks
with a characteristic time scale $ t_{_{1/2}} \sim t_0 / A_{max} $,
where $ A_{max} \gg 1 $ is the peak magnification factor.  The two
almost equally bright
images are separated by $ \sim 2'' ~ (M/10^6 ~ M_{\odot} )^{1/2} $, and
they rotate very rapidly around the lens with the relative proper
motion enhanced by a factor $ \sim 2 A_{max} $.  The same events
will offer an opportunity to study spectroscopically stars which
are normally far too faint to be reached.

\vskip 0.5cm
{\bf Key words}: {\it black holes -- dark matter -- gravitational lensing}

\vskip 0.5cm

The extension of the search for MACHOs to masses as large as
$ 10^3 - 10^6 ~ M_{\odot} $ (cf. Lacey and Ostriker 1985)
is difficult because the characteristic
time scales of the microlensing events caused by them are very long,
$ t_0 \sim 7 - 200 $ years, while the current searches are most sensitive
to $ 10 \leq t_0 \leq 100 $ days (Udalski et al. 1994a, Alcock et al. 1995).
Even though the upper limit will gradually
increase as the searches continue, it is clear that $ t_0 \sim 100 $ years
is out of reach.  In principle, one may identify such long duration
events trying to uncover the very small photometric variability due to the
parallactic effect (Gould 1992), but the required photometric accuracy is
$ \sim 1\% $, and that is very hard to achieve in a very dense
stellar field for millions of stars.
However, if the impact parameter in the lensing geometry
is very small, and the corresponding peak magnification is very large,
$ A_{max} \gg 1 $, than the time scale on which the magnification
changes between $ A_{max} $ and $ A_{max}/2 $ is only
$ t_{_{1/2}} \sim t_0/A_{max} $, i.e. it may be detectable
in a few years with an ordinary photometric accuracy.

The obvious problem is that the very high magnification events are very rare.
However, as they
bring up into visibility a huge number of very faint stars which
are normally undetectable, the rate of such events may be relatively high.
The aim of this paper is to study the feasibility of a
search for very massive compact objects with very
high magnification microlensing events of the stars in nearby galaxies.
Some other problems related to microlensing of stars which are
normally below the detection threshold were studied by Colley (1995),
Crotts (1992), Gould (1995) and Nemiroff (1994).

A gravitational lensing system made of a
point mass located at a distance $ D_d $ lensing a source star located
at a distance $ D_s $ has an Einstein ring radius given as
$$
\varphi _{_E} = 0.''9 ~
\left( { M \over 10^6 ~ M_{\odot} } ~ { 10 ~ kpc \over D_d } \right) ^{1/2}
\left( 1 - { D_d \over D_s } \right) ^{1/2} ,
\eqno(1)
$$
(cf. Paczy\'nski 1986).
If the relative transverse velocity of the lensing mass is $ V $ then
the characteristic time scale of the lensing event is
$$
t_0 = { \varphi _{_E} \over \dot \varphi } = 214 ~ yr ~
\left( { M \over 10^6 ~ M_{\odot} } ~ { 10 ~ kpc \over D_d } \right) ^{1/2}
\left( 1 - { D_d \over D_s } \right) ^{1/2}
\left( { 200 ~ km ~ s^{-1} \over V } \right) ,
\eqno(2)
$$
where $ \dot \varphi = V/D_d $ is the proper motion of the lens.

The magnifications of the two images formed by the lens are
$$
A_{1,2} = { u^2 +2 \over 2 u ( u^2 + 4 )^{1/2} } \pm { 1 \over 2 } ,
{}~~~~~~ A = A_1 + A_2 = { u^2 +2 \over u ( u^2 + 4 )^{1/2} },
{}~~~~~~ A_1 - A_2 = 1 , \eqno(3)
$$
and their angular distances from the lensing mass are given as
$$
r_{_{1,2}} = { 1 \over 2 } \left[ ( u^2 + 4 )^{1/2} \pm u \right],  \eqno(4)
$$
where $ u $ is the angular distance between the source and the lens
divided by the Einstein ring radius.
Let the source move along the `x' axis with impact parameter $ p $.
The coordinates of the source are given as
$$
x = t / t_0 , ~~~~~~~~~~~~ y = p = const ,  \eqno(5)
$$
where $ t_0 $ is the characteristic time scale of the lensing event
defined as the Einstein ring radius divided by the source velocity,
and the coordinates $ (x,y) $ are also in units of the Einstein
ring radius.  We adopt $ t = 0 $ for the time of the peak magnification.
Of course, we have
$$
u^2 = x^2 + y^2 .  \eqno(6)
$$
The position angles of the two images with respect to the lensing mass are
given as
$$
\tan \theta _1 = { x \over p } , ~~~~~~~ \theta _2 = \theta _1 + \pi .
\eqno(7)
$$

Let us consider now the case of a very high magnification, i.e. $ p \ll 1 $.
We may expand the equations (3) and (4) and retain the leading terms to obtain:
$$
A = { 1 \over u } = { 1 \over ( p^2 + x^2 )^{1/2} } =
A_{max} ~ \left[ 1 + \left( { x \over p } \right) ^2 \right] ^{-1/2} ,
\eqno(8)
$$
$$
r_{_{1,2}} = 1 \pm { u \over 2 } .  \eqno(9)
$$
The characteristic time scale for the brightness variation, $ t_{_{1/2}} $
is the time it takes the magnification to change from
$ A_{max} = 1/p $ to $ A_{1/2} \equiv A_{max}/2 $, i.e. by 0.76 magnitude.
It follows from the eqs. (8) and (5) that
$$
t_{_{1/2}} = t_0 ~ x_{_{1/2}} = t_0 ~ p ~ \sqrt{3} =
\sqrt{3} ~ { t_0 \over A_{max} } , \eqno(10)
$$
i.e. $ t_{_{1/2}} $ is much shorter than $ t_0 $.

While the magnification changes from $ A_{max}/2 $ to $ A_{max} $ and
back to $ A_{max}/2 $ the position angle of the first image varies
from $ \theta _1 = - \pi /3 $ through $ \theta _1 = 0 $ to
$ \theta _1 = \pi /3 $, and the rate of change is
$$
{ d \theta _1 \over dt } = { \sqrt{3} \over t_{_{1/2}} }
\left[ 1 + 3 \left( { t \over t_{_{1/2}} } \right) ^2 \right] ^{-1} =
{ \sqrt{3} \over t_{_{1/2}} } \left( { A \over A_{max} } \right) ^2 =
{ A^2 \over t_0 A_{max} } .  \eqno(11)
$$

We have the following approximate picture of a very high magnification
event with a very small impact parameter, $ p \ll 1 $.  The
two images circle the lens, the primary just outside the Einstein
ring, while the secondary just inside it (cf. eq. 9).  Their near circular
motion first accelerates while the intensity increases, and later
decelerates past the light maximum (cf. eq. 11).  The two images are
almost equally bright.  Their angular separation stays very close to the
diameter of the Einstein ring, which is approximately given as
$$
2 ~ \varphi _{_E} = 1.''8 ~
\left( { M \over 10^6 ~ M_{\odot} } ~ { 10 ~ kpc \over D_d } \right) ^{1/2} ,
\eqno(12)
$$
(cf. eq. 1) where we assume that the source star is very far away,
i.e. $ D_d / D_s \ll 1 $.  Note, that if a supermassive lens is in the
halo of the external galaxy, like M31, then the image separation
as given with the eq. (1) is very small, as $ D_d / D_s \approx 1 $.

The real time discovery of the very high magnification
events of stars in other galaxies will be possible when the
OGLE's Early Warning System (Udalski et al. 1994b) is extended
to `new' stars, not present on the templates
currently used for the rapid star identification.  Once the candidate
event is located the highest resolution astrometry should be attempted
in order to resolve the double image.  Current ground based infrared
imaging achieves the resolution of $ \sim 0.''2 $ (Eckart et al. 1993).
Even higher resolution is possible with the Hubble Space Telescope.
Still higher resolution might be achieved with adaptive optics and imaging
interferometry (cf. Shao and Colavita 1992, Beckers 1993, Robertson and
Tango 1994, Roddier 1995, Shao 1995, and references
therein).  The resolution of $ \sim 0.''01 $ would correspond to a
lensing mass of $ \sim 25 ~ M_{\odot} $, and $ t_0 \sim 1 $
year, i.e. the range directly
accessible to standard microlensing searches among the stars which are
detectable at all time.  Note, that if resolved,
the very high magnification double image offers unmistakable evidence
that it is due to microlensing: the two images rotate rapidly around
the lens, with the peak relative proper motion given as
$$
2 ~ \dot \varphi _{_A} = 2 ~ \varphi _{_E} ~ \dot \theta _1 =
2 ~ { \varphi _{_E} \over t_0 } ~ { A^2 \over A_{max} } =
\dot \varphi ~ { 2 A^2 \over A_{max} } , \eqno(13)
$$
where $ \dot \varphi $ is the proper motion of the lensing object.  We
assume the lensed star to be so far away that its proper motion is negligible.

Direct detection of a double image would not only provide a direct proof
of the microlensing event, it would also provide an estimate of the lens
mass through eq. (12).  In case of very high magnification events
an alternative possibility to estimate the lens mass is available when the
source star of a known size is resolved by the lens
(Nemiroff and Wickramasinghe 1994, Witt and Mao 1994, Witt 1995).
In the very rare case when the source is resolved by the lens, and at
the same time the image is resolved directly, the image would
appear to be not just double, but ring-like.

The obvious problem with the proposed method of extending the microlensing
search to supermassive objects is the low probability of the very high
magnification events.  The standard optical depth $ \tau $ is defined as
the probability that any particular star is magnified by a factor larger than
$ 3 / \sqrt{5}
\approx 1.34 $ (corresponding to $ u = 1 $ in eq. 3).  The probability
that at any given time a particular star is magnified by a factor
$ A \gg 1 $ is equal to $ \tau / A^2 $.  However, as the duration of
each very high magnification event, $ t_{_{1/2}} $, is shorter than $ t_0 $
by a factor $ A $, the rate of the very high magnification events is
lower than ordinary events only by a factor A.  Therefore, if the
luminosity function below the detection threshold is inversely proportional
to the stellar luminosity, i.e. the number of faint stars is larger
in inverse proportion to their luminosity, then the very high magnification
event rate is independent of their magnification factor.
However, for a given lensing mass the effective duration of those
events, $ t_{_{1/2}} $, is inversely proportional to $ A_{max} $ (cf. eq. 10).
This implies that a larger number of  photometric measurements is required
to reliably detect those events.

Our conclusion is: it should be possible to either detect supermassive
objects in the galactic halo if those objects exist, or to put stringent
upper limits on their number density, within a sensibly long observing
project, like 10 years.  However, it will be necessary to make frequent
photometric measurements
to detect the short duration peaks of very the high magnification events
of faint stars which are normally below the detection threshold.
It will be very helpful to have real time identification of the
events with a new Early Warning System.
The same project will make it possible to obtain spectra of the lensed
stars, which are normally too faint to be reached.

\vskip 1.0cm

{\bf Acknowledgements.} It is a pleasure to acknowledge comments
by Dr. Joachim Wambsganss.  This project was supported with the
NSF grants AST 92-16494 and  AST 93-13620.

\vskip 0.5cm
\centerline{REFERENCES}
\vskip 0.5cm

\wc{Alcock, C. et al. 1995, {\it Phys. Rev. Letters}, {\bf 74}, 2867.  \hfill}

\wc{Beckers, J. M. 1993, {\it Ann. Rev. Astron. Ap.}, {\bf 31}, 13. \hfill}

\wc{Colley, W.N.  1995, {\it A. J.}, {\bf 109}, 440.  \hfill}

\wc{Crotts, A.P.S. 1992, {\it Ap. J.}, {\bf 399}, L43.  \hfill}

\wc{Eckart, A. et al. 1993, {\it Astrophys. J. Letters}, {\bf 407}, L77.
\hfill}

\wc{Gould, A. 1992, {\it Ap. J.}, {\bf 392}, 442.  \hfill}

\wc{Gould, A. 1995, {\it Ap. J.}, {\bf 435}, 573.  \hfill}

\wc{Lacey, C. G., and Ostriker, J. P. 1985, {\it Astrophys. J.},
{\bf 299}, 633.  \hfill}

\wc{Nemiroff, R. J. 1994, {\it Astrophys. J.}, {\bf 435}, 682.  \hfill}

\wc{Nemiroff, R. J., and Wickramasinghe, W. A. D. T.  1994,
{\it Astrophys. J.}, {\bf 424}, L21.  \hfill}

\wc{Paczy\'nski, B. 1986, {\it Astrophys. J.}, {\bf 304}, 1. \hfill}

\wc{Robertson, J. G., and Tango, W. J. (Editors) 1994, IAU Symp. No. 158:
{\it Very High Angular Resolution Imaging}.  \hfill}

\wc{Roddier, F. 1995, {\it Astrophys. Space Sci.}, {\bf 223}, 109.  \hfill}

\wc{Shao, N., and Colavita, M. M. 1992, {\it Ann. Rev. Astron. Ap.}, {\bf 30},
457.  \hfill}

\wc{Shao, M. 1995, {\it Astrophys. Space Sci.}, {\bf 223}, 119.  \hfill}

\wc{Udalski, A. Szyma\'nski, M., Stanek, K.Z., Kaluzny, J., Kubiak, M.,
Mateo, M., Krzemi\'nski W., Paczy\'nski, B., and Venkat, R.  1994a,
{\it Acta Astron.}, {\bf 44}, 165. \hfill}

\wc{Udalski, A. Szyma\'nski, M., Kaluzny, J., Kubiak, M.,
Mateo, M., Krzemi\'nski W., and Paczy\'nski, B. 1994b,
{\it Acta Astron.}, {\bf 44}, 165. \hfill}

\wc{Witt, H. J. 1995, {\it Astrophys. J.}, in press.  \hfill}

\wc{Witt, H. J., and Mao, S.  1994, {\it Astrophys. J.}, {\bf 430}, 505.
\hfill}

\vskip 2.0cm
\centerline{This paper has been submitted to Acta Astronomica on May 4, 1995}

\vfill
\end
\bye